\documentstyle[11pt]{article}
\begin{document}
\title{The Role of the Future in Quantum Theory}
\author{Paul Sommers\\High Energy Astrophysics Institute\\
Department of Physics\\University of Utah\\Salt Lake City, Utah 84112}
\maketitle
\vspace{1cm}
\begin{quote} \centering
gr-qc/9404022

\vspace{.5cm}
ABSTRACT
\end{quote}
\vspace{1cm}
\baselineskip=6mm

	Interpretation problems are eliminated from quantum theory by
picturing a quantum history as having been sampled from a
probability distribution over the set of histories which are
permitted by all relevant boundary conditions.  In laboratory
physics, the final measurement plays a crucial role in defining
the set of allowed histories by constraining the final state to
be an eigenstate of the measurement operator.  For the universe
itself, a final boundary condition can play a similar role.
Together with the special big bang initial state, the final
constraint may ensure that the universe admits a classical
description in the asymptotic future.  Acknowledging the role
of a future constraint dispels the mysteries of quantum theory
without any amendment to the theory itself.

\vspace{1cm}
\section{Introduction}

	In a typical experiment, the initial state of a quantum
system is fixed by its preparation and the final state is
measured.  The eigenstates of the measurement operator constitute
the set of allowed final states.  The formalism of quantum
theory yields a probability distribution on the set of allowed
final states, and experiments consistently verify the validity of
the formalism.  The relationship between experiments of that type
and the formalism is clear.  Problems of interpretation do not
arise if every quantum system is subject to the type of future
constraint which is imposed by a final measurement.  This paper
considers how a similar final constraint might operate on a
closed system like the 4-dimensional universe.  There is no
external apparatus to perform a final measurement, so the final
constraint must be considered as a boundary condition.

	The hypothesis of a final constraint offers an approach
to quantum theory which is free of interpretation problems and is
consistent with any conceivable test.  Any condition which is
satisfied by the actual final state of the universe could be an
operative boundary condition.  No experiment in this universe
can determine whether or not the satisfied final condition is
an imposed boundary condition.  The possibility of a future
constraint is here taken seriously because it cannot be excluded
and because it renders quantum theory understandable.  Objective
reality is maintained without a ``many worlds'' interpretation
\cite{s1}.  State reductions occur without amending quantum
theory with an ad hoc R-process
\cite{s2}.  The limitation on causality is no different than for
a laboratory quantum system whose final state is constrained by
an external measurement apparatus.

	Speculation about the actual nature of the final
constraint is offered in sections 3-5.  The simplest hypothesis
is that the final state is required to be sampled from a special
set of basis states, exactly as if an external measurement were
made.  However, due to the highly special initial big bang state,
it may be that the final constraint need not be so specific as to
select a unique final basis.  Many different physically
reasonable final bases may define the same probability
distribution over a common set of physically equivalent quantum
histories.  In the context of quantum gravity, the final
constraint may be tantamount to requiring that a geometric
description should pertain to the universe, at least for
asymptotically late times.

	The need for a future constraint is suggested by the
Einstein-Podolsky-Rosen (EPR) phenomenon \cite{r2}.  As presented
by Bohm \cite{r3}, the EPR effect concerns the decay of a spin-0
system into two spinning particles.  Measurement of the spins of
both particles with respect to a common spin axis always yields
equal-but-opposite spin values (as required by angular momentum
conservation).  The perfect anti-correlation holds regardless of
the choice of spin axis for the measurement, and even if the spin
axis is selected by a random process after the two spinning
particles are well separated \cite{r4}.  Further experimental
results verify
\cite{r4}, as expected from the non-commutation of spin operators
associated with different spin axes, that the two particles
cannot carry prepared answers to all possible spin tests
\cite{r5,r6}.  That being the case, how can they be certain to
give opposite answers to every possible test?  No satisfactory
answer is available in the context of future-blind evolution.
There is nothing paradoxical about the EPR phenomenon, however,
if one regards an entire quantum history as having been selected
as a single entity.  Each allowed history necessarily satisfies
all conservation laws.  There is no allowed history in which both
particles are measured with positive spin components along the
z-axis, for example, and so that situation never occurs.  The
particles do not evolve without reference to the measurements in
their future.  The entire quantum history is a single entity.
Even though the orientation of the future spin measurement might
not be knowable at the time of the spin-0 decay, each of the
allowed histories nevertheless includes a specific measurement
orientation, and both particles might be regarded as having been
emitted in eigenstates of the corresponding spin operator.  There
is no paradox in the EPR phenomenon if one accepts that the
emitted particles are conditioned by their future in this sense.

	This approach requires a clear understanding for the term
``history'' in quantum theory.  It is not simply the Schroedinger
evolution of the initial state, since that evolved state is
generally a grotesque superposition \cite{g1} of physically
interpretable states at late times.  The quantum history of the
universe includes a rich sequence of reductions to physically
interpretable states.  The occurrence of those state reductions
is here attributed to a final constraint condition together with
the special initial state.  Before considering specific ideas
about the final constraint and what is meant by a ``quantum
history,'' however, it will first be argued that there is nothing
unacceptable about a future constraint.

\section{Boundary conditions and causality}

	Quantum theory does not favor initial boundary conditions
over final boundary conditions or mixed boundary conditions.  The
theory itself is time-symmetric.  The rules of quantum theory
assign a transition amplitude to any pair of initial and final
states.  A transition amplitude is time-symmetric in the
following sense: If S is the transition amplitude for state
$\psi$ to become state $\phi$, then S is also the amplitude for
time-reversed $\phi$ to become time-reversed $\psi$ (with all
interaction laws operating in their time-reversed form).

	Consider the specific example of a photon which passes
through a double-slit diffractor and is absorbed on a subsequent
screen.  There is a particular amplitude S for its absorption at
a particular point P on the screen.  Now imagine the time reverse
of absorption at P, so a photon of the same frequency is emitted
from P toward the diffractor.  Such a photon can go off in many
directions from the diffractor.  Most likely it will be absorbed
somewhere on a laboratory wall.  However, the amplitude for its
absorption in the lamp is equal to the amplitude S for the photon
emitted from the lamp to be absorbed at P.

	Quantum theory is time-symmetric in this narrow sense of
time symmetry for transition amplitudes.  It is not the time
symmetry which pertains to deterministic classical physics where
each initial state identifies a unique final state and vice
versa.

	Time symmetry of quantum theory must also be clearly
distinguished from time symmetry (or not) of particular
interaction laws.  If all interaction laws were time-symmetric,
then the amplitude for going from state $\psi$ to state $\phi$
would be the same as the amplitude for time-reversed $\phi$ going
to time-reversed $\psi$, period.  If the interaction laws are not
time symmetric, however, then the amplitude for $\psi$ going to
$\phi$ may have two different values, depending on the
orientation of time.  The amplitude for $\psi$ going to $\phi$
with one orientation of time is equal to the amplitude for
time-reversed $\phi$ going to time-reversed $\psi$ with the
opposite time orientation.  The narrow meaning of time symmetry
in quantum theory therefore applies also in the presence of
interaction laws which are not time-symmetric.

	A universe with a final boundary condition can be
imagined as easily as a universe with an initial boundary
condition.  It is customary to assume that the universe has
evolved from the low entropy state in the past.  In the context
of deterministic classical physics, however, it is not possible
to reject the interpretation that the universe is evolving
backward in time from the very special final state which could be
obtained by the classical evolution of the low entropy initial
state.  In that perverse (but allowed) interpretation, causes are
ascribed to future events instead of past events, and one may
perversely maintain that no system is influenced by events
outside its causal future.  The customary time orientation of
causality derives from asymmetric boundary conditions.  Indeed,
were the low-entropy state an imposed final state rather than an
imposed initial state, the customary and perverse notions of
causality would be interchanged.  The theory itself is time
symmetric and does not favor any particular boundary conditions
or time orientation for causality.

	In quantum physics, the initial state of a system does
not determine its future.  Schroedinger evolution of the initial
state yields only an amplitude for transition to any future
state.  In laboratory experiments, a final measurement defines an
orthonormal set of allowed final states for the system.  The
squared norms of the transition amplitudes to those measurement
eigenstates constitute a properly normalized probability
distribution.  One may identify the set of allowed quantum
histories with the set of final measurement eigenstates, so there
is a probability distribution over the set of allowed quantum
histories.  One can picture the final state (and hence the
quantum history) as being sampled from that probability
distribution.  For this same picture to apply to the quantum
universe, some final constraint must similarly restrict the
allowed histories to a set over which there is a normalized
probability distribution.  A random sampling from that
probability distribution can then give rise to the quantum
history of the universe.

	The hypothesis that the universe is subject to a future
constraint raises questions about causality.  The standard point
of view is that boundary conditions apply only in the past, and
the future is completely unconstrained.  Stemming from this, the
strong principle of causality asserts that a system cannot be
influenced by events or conditions which lie outside its causal
past.  The existence of a future constraint certainly conflicts
with these strong ideas about causality.  On the other hand, the
final constraint might be analogous to the effect of a final
measurement by an external apparatus.  How much damage does that
do to causality?  The EPR experiment can be used to study the
question.  Whenever the spins of two particles produced by the
decay of a spin-0 state are tested with respect to a common spin
axis, they behave as if they were produced in eigenstates of that
spin operator.  For example, the two particles seem to be emitted
in x-spin eigenstates if x-spin measurements are subsequently
performed on both; they seem to be emitted in y-spin eigenstates
if y-spin measurements are subsequently performed on both; etc.
It does not matter how or when the measurement axis is selected.
The choice of measurement axis, regardless of how or when it is
made, limits the set of possible histories, and the entire
quantum history is sampled as a single entity.  If one insists on
picturing systems as evolving in time, however, it appears that
the particles are influenced at their event of production by the
measurements performed at some later time.  This violates the
strong notion of causality cited above.  Nevertheless, there is
no way to extract the information about the orientation of the
future spin measurement axis.  The particles' spin states cannot
be tested without disrupting the influence of the future
measurements.  The influence from the future therefore does not
constitute a signal propagating backward in time.  Note also that
quantum theory respects the light cone structure of relativity in
the sense that a single particle cannot be localized at two
spacelike-separated events, so particles do not transport signals
faster than the speed of light.  If backward-in-time signaling
does not occur, then the practical notion of causality can be
maintained: ``The maximum signal speed is the speed of light.''
As noted previously, causality is not intrinsic to the theory
itself, but derives from boundary conditions.  With an imposed
low-entropy {\em final} state, one would interpret all signals as
propagating backward in time.  With fully time-symmetric boundary
conditions, causality would not be a useful notion at all.  Due
to asymmetric boundary conditions, causality {\em is} a practical
notion in the actual universe, but the strongest formulation of
the principle may not be tenable.

\section{Quantum history as beginning-to-end transition}

	There is no interpretation problem associated with the
EPR phenomenon if a quantum history is sampled from a probability
distribution over an allowed set of histories each of which
satisfies all conservation laws.  This approach requires a
specific meaning for the term ``quantum history.''  In laboratory
physics, the set of allowed quantum histories is governed by the
measurement chosen to determine the final state as well as the
system's preparation which determines the initial state.  Given a
fixed initial state $\psi_{0}$, the allowed quantum histories may
be identified with transitions to the possible eigenstates
$\psi_{k}$ of the final measurement operator.  By virtue of that
identification, the quantum formalism provides a probability for
each allowed history.  The probability for the history linking
$\psi_{0}$ to $\psi_{k}$ is $|<\psi_{k}|H|\psi_{0}>|^{2}$, where
$H$ maps the initial state to the final time via Schroedinger
evolution (or the complex conjugate operator maps the final state
to the initial time) and the bracket denotes the Hilbert space
inner product.  Experiments consistently verify the validity of
those probabilities.  This paradigm pertains to all laboratory
quantum physics.  The operative constraints and/or boundary
conditions define a set of allowed quantum histories.  Only one
actual history occurs.  The quantum formalism concerns the
probability distribution for any set of allowed histories.

	For a closed system like the 4-dimensional universe,
there is no external measuring apparatus to impose a special set
of basis states from which the final state must be drawn.
However, by supposing that a final boundary condition similarly
requires the final state to be drawn from some special set of
final basis states, the same notion of a quantum history as a
beginning-to-end transition can pertain as well to the 4-dimensional
universe.  A probability amplitude is then associated with each
allowed history, and the quantum history of the universe may be
regarded as having been sampled from that probability
distribution.  Only one of the allowed histories actually occurs.

	A quantum history is therefore characterized by a
transition from the initial state to an allowed final state.  In
a laboratory scattering experiment, the final measured state may
permit a reconstruction of particle worldlines, so the quantum
history determines a classical history.  That is not always the
case, however, as illustrated by a photon which passes through a
double-slit diffractor and is detected on a subsequent screen.
Simply measuring the final position of the photon does not
determine which of the possible classical worldlines it followed.
The quantum history does not admit a unique classical
description.  The situation is different if a measurement of the
photon is made at one of the slits.  In that case the quantum
history is defined by a final state which contains a record of
which slit the photon went through, and the photon's unique
worldline is implicit in the quantum history.  Now suppose the
measurement at the slit is reversible in the sense of Wigner
\cite{r7}.  If the measurement is actually reversed, then there
is no record of its outcome in the final state, and the two
possible worldlines interfere coherently.  Conversely, if the
measurement interaction is not reversed, then the final state
does contain a record of the photon's position in one or the
other slit, and a unique worldline is determined.  The measurement
interaction at a slit does not by itself localize the particle,
since the measurement could subsequently be reversed.  Whether or
not the particle is localized at a slit is a property of the
whole quantum history, because it is determined by the final
state.

	The characterization of a quantum history by its initial
and final states clarifies the concepts of ``successful
measurement'' and ``state reduction.''  Let $\psi_{0}$ and
$\psi_{f}$ denote the initial and final states of a quantum
history.  Suppose $\phi$ is a state at some intermediate time
with the property that $$<\phi|\phi'>=0 \hspace{5mm} \Rightarrow
\hspace{5mm} <\psi_{f}|H_{2}|\phi'><\phi'|H_{1}|\psi_{0}>=0,
\hspace{2 cm} (1)$$ where $H_{1}$ is the unitary Schroedinger
mapping of states from the initial time to the intermediate time
and $H_{2}$ maps states from the intermediate time to the final
time.  If $\phi$ is an eigenstate of some measurement interaction
between subsystems, for example, then it represents a successful
measurement.  Condition (1) means that none of the other
(orthogonal) eigenstates $\phi'$ is both accessible from the
initial state and compatible with the final state.  It also means
that the amplitude for transition from $\psi_{0}$ to $\psi_{f}$
is not reduced by projection onto $\phi$ at the intermediate
time:
$$<\psi_{f}|H_{2}H_{1}|\psi_{0}>=<\psi{_f}|H_{2}|\phi><\phi|H_{1}|\psi_{0}>.
\hspace{3cm} (2)$$
In the history defined by $\psi_{0}$ and
$\psi_{f}$, therefore, the outcome of the measurement is
definitely $\phi$, and one says that the system is reduced to the
state $\phi$ at the intermediate time.

	The Schroedinger cat gedanken experiment \cite{r8}
provides a dramatization of these ideas.  A cat is enclosed in a
box along with a quantum subsystem which may or may not decay in
a prescribed time interval.  Decay of the subsystem would cause a
vial of poison to be broken so the cat would die.  The box and
its contents constitute a quantum system starting in a pure
state.  When the box is opened after the prescribed time
interval, the Schroedinger-evolved pure state entails a
superposition of live cat and dead cat.  Does any state reduction
to either the live-cat state or the dead-cat state occur?  The
answer lies in the final state, because that is what defines the
quantum history.  Part of the problem in the Scroedinger cat
experiment is deciding what is meant by the final state, since
there is a sequence of correlated outcomes.  If the atom decays,
then the vial is broken, then the cat dies, then an intelligent
observer recognizes that the cat is dead, etc.  Is there any
point beyond which superposition with the alternative outcome is
impossible?  In principle, the answer is No.  The
resolution must be sought in the final state $|\Psi_{f}>$ of the
universe itself.  If no permanent record exists of whether the
cat was alive or dead when the box was opened, then the final
state can be reached through either path, and neither classical
description by itself pertains to the quantum history.  However,
the interference of different macroscopic states does not
normally occur.  It is therefore natural to assume that the final
state must contain some record of whether the cat was dead or
alive when the box was opened.  Symbolically, either
$<\Psi_{f}|H_{2}|\Psi_{live}>=0$ or
$<\Psi_{f}|H_{2}|\Psi_{dead}>=0$ must be true, where $H_{2}$ is
the unitary mapping from the time of box opening to the time of
the final state.  Then, since the initial state is mapped by
$H_{1}$ to a superposition of just $|\Psi_{live}>$ and
$|\Psi_{dead}>$, one or the other of those states satisfies the
conditions (on $\phi$ in equation 1) for a state reduction at the
time of the box opening.  The state reduction means the quantum
history does {\em not} entail a superposition of live cat with
dead cat.

	It remains to explain {\em why} the final state of the
universe should be incompatible either with the live cat state or
else with the dead cat state.  After all, if $|\Psi_{1}>$ is a
final state which is compatible only with the live cat state and
$|\Psi_{2}>$ is another final state which is compatible only with
the dead cat, then any superposition of them is a valid quantum
state.  The quantum history defined by such a superposition final
state would not include a state reduction to a live cat or dead
cat at the intermediate time.  What is it that prohibits a
superposition final state of that type?  There is nothing in
quantum theory itself to exclude superpositions of
$|\Psi_{1}>$ and $|\Psi_{2}>$.  If quantum theory is not to be
amended with additional structure, then the exclusion must be
attributed to a boundary condition.  The states $|\Psi_{1}>$ and
$|\Psi_{2}>$ may separately satisfy an imposed boundary condition
which no superposition of them could satisfy.  The following two
sections address this issue of what constitutes a physically
plausible boundary condition, i.e. why an allowed final state is
extremely unlikely to be accessible from both the dead cat state
and the live cat state.

\section{Toy cosmology}

	In a simplified cosmology without gravitation or other
long-range interactions, one can imagine that at $t=0$ there is
only an extremely massive particle at the origin of Euclidean
space.  There are no other particles or radiation.  The massive
particle is allowed to decay, and a stochastic cascade of
inelastic collisions, decays, and radiative processes distributes
the energy among more and more particles.  In the asymptotic
future, there are a vast number of low-energy free particles.  The
spacetime picture of a cascade is a network of intersecting
particle worldlines.  The network is similar in some respects to a
giant tree stemming from a single root, with a multitude of
smaller and smaller branches.  Its leaves correspond to classical
free particle worldlines in the asymptotic future.  Because of
the stochastic nature of the decays and interactions, the
original massive particle does not determine the cascade.  (Many
different trees may have an identical root.)

	Two or more particle networks which are identical in the
asymptotic future are here regarded as a single cascade.  The
different intermediate developments interfere coherently if the
asymptotic cascade contains no record of one or the other
development having occurred at the intermediate times.  A cascade
may be defined as a class of semi-classical (stationary-phase)
Feynman paths \cite{g2} which are represented in the asymptotic
future by the same classical phase space trajectory.

	Each cascade can be characterized by some
information from its asymptotic classical phase space trajectory.
The set of asymptotic trajectories which can result from the
special initial state is extremely restricted, so a relatively
small amount of information about the final trajectory suffices
to identify an individual cascade.  In particular, it is
plausible that the amount of information which specifies a final
{\em quantum} state, together with the special initial
conditions, is adequate to identify an individual cascade.  For
example, suppose all particle positions are specified arbitrarily
at some late time.  Can a momentum be assigned to each
final particle so that the final state can be evolved backward
semi-classically to the special concentration in a single mass at
the origin of Euclidean space at $t=0$?  In general, the answer
is No.  Only special final particle configurations at the late
time can be attained from the various possible cascades.
Moreover, a final configuration which can be attained from a
cascade comes from (in general) at most one such cascade.  It is
highly exceptional for two or more distinct cascades to yield
identical particle positions at some late time.  A cascade can be
determined, therefore, by the initial condition together with
final particle positions.

	The reason why each cascade can be identified by
relatively little information about its asymptotic trajectory is
that the cascade develops by virtue of particle decays and also
inelastic collisions with more than two outgoing particles.  The
time reverse of such processes requires extremely special
kinematic conditions.  The time reverse of particle decay is
fusion without radiation, a phenomenon which is too special to be
observed in nature.  Similarly, inelastic collisions which reduce
the number of particles can only occur if kinematic conditions
are finely tuned.  The semi-classical backward evolutions of a
generic asymptotic phase space trajectory would afford few
opportunities for such processes, so it could not coalesce into
the single massive particle at $t=0$.  Consequently, a cascade
which develops from the single massive particle can be
characterized at some late time by particle information which
could be provided by a final quantum state.

	The above assertions are supported quantitatively by
comparing the degrees of freedom in such a cascade with the
amount of information provided by a final quantum state.  Suppose
the cascade leads to a total of $n$ particles in the final state.
A final quantum state can be specified by $3n$ parameters,
which might be the final coordinates of all the particles, or the
components of their final momenta, etc.  The semi-classical
cascade can be represented by a network of intersecting particle
worldlines in 4 dimensions.  There are a countable number of
topologically distinguishable networks which start from a given
worldline and end with $n$ particles.  For each topology, the
various possible networks can be parametrized by fewer than $3n$
numbers.  Consider, for example, what happens when a particle
decays into two.  There is one parameter's freedom in specifying
when the decay occurs, and two parameters to specify the
direction of one of the outgoing particles.  With fixed masses
for the outgoing particles, the 4-momenta of the two outgoing
particles are determined by energy-momentum conservation once the
first direction is specified.  So there are 3 parameters to
specify this decay which increases the number of particles
by one.  Similarly, if the number of particles increases by $m$
as the result of an inelastic collision of two particles, there
are $3m$ additional parameters as a result.  (To arrange for the
collision of two particles which would otherwise have
unconstrained momenta, one of them loses two degrees of freedom.
Conserving energy-momentum with particles of fixed mass leaves
$3(m+2)-4$ degrees of freedom in the 4-momenta of the $m+2$
outgoing particles.  This compensates for the two degrees lost to
arrange the collision and also adds $3m$ additional parameters
associated with this production of $m$ additional particles.)
The final state occurs after an increase of $n-1$ particles from
the initial single mass.  The freedom in the cascade development
is parametrized by $3(n-1)$ parameters.  In the $3n$-parameter
space of allowed final quantum states, therefore, only a special
subset admit a semi-classical particle cascade interpretation.

	In any of these toy universes, entropy increases from 0
(if there is only one state for the initial massive particle) to
a very large value as the energy is distributed among more and
more particles.  Due to the stochastic nature of the
interactions, a specification of the cascade at some intermediate
time $t$ is insufficient for reliably predicting the future.  On
the other hand, if a quantum state at time $t$ is compatible with
some cascade from the initial state, then the cascade up to that
time is (in general) uniquely determined.  As a result, past
history {\em can} be confidently inferred (using knowledge of the
initial state).

	It is natural to think of a cascade as a quantum history.
This is an apparent departure from the earlier definition of a
quantum history as a beginning-to-end transition.  However,
within the context provided by the highly special initial state
of the toy cosmology, each cascade corresponds to a special class
of beginning-to-end transitions, and this correspondence endows
the set of cascades with a probability distribution.  It is thus
possible to regard a toy universe as a cascade sampled from that
probability distribution.

	Each cascade has a unique asymptotic classical phase
space trajectory.  As a result, it identifies a unique final
position eigenstate, or a unique final momentum eigenstate, etc.
Moreover, due to the special nature of the initial state, a final
eigenstate of position or momentum (etc.) is almost always
compatible with at most one cascade.  Consider the final momentum
eigenstates.  There is a subset for which each eigenstate
corresponds to at least one cascade.  Those are the eigenstates
for which there is a non-negligible transition amplitude.  (The
others have no stationary-phase Feynman path to contribute to
their amplitudes in the sum-over-paths method.)  Among those
final momentum eigenstates which correspond to cascades, there
could be ones which are compatible with more than one cascade.
(That requires another accessible asymptotic phase space
trajectory with identical particle momenta.)  Those exceptional
eigenstates constitute a set of measure zero, however.  The
transition probabilities therefore give a normalized probability
distribution over the set of those momentum eigenstates which
correspond to unique cascades.  This probability distribution can
then be considered as pertaining to the set of cascades
themselves.  The same probability distribution results from the
correspondence of cascades with final position eigenstates (or
some other physically reasonable basis).  The probability
pertains to the cascade, not just a particular final state in
correspondence with it.

	The special initial state is responsible for the
existence of a probability distribution over the set of cascades.
Were the initial state a typical $n$-particle position
eigenstate, for example, no final $n$-particle position
eigenstate and no final momentum eigenstate would correspond to a
unique cascade.  (The amplitude for transition to each such
accessible eigenstate would include contributions from multiple
stationary-phase Feynman paths with disparate asymptotic
classical phase space trajectories.)  The almost 1-to-1
correspondence between cascades and the accessible basis states
for physically reasonable final bases is a remarkable consequence
of the special initial state in the toy cosmology.  (The
correspondence may be the property which characterizes a
``physically reasonable'' final basis, a property which no basis
would have without the special initial state.)

	A cascade identifies many different beginning-to-end
transitions since it identifies a final momentum eigenstate,
final position eigenstate, etc.  All beginning-to-end transitions
which correspond uniquely to the same cascade should be regarded
as physically equivalent.  A transition to a final position
eigenstate might imply a sequence of position eigenstate
reductions at intermediate times (as defined by condition 1),
whereas an equivalent transition (via the same cascade) to a
final momentum eigenstate might imply a sequence of reductions to
momentum eigenstates at intermediate times.  Both sets of state
reductions pertain to the same cascade.  In a cascade history,
the system can be simultaneously reduced to a position eigenstate
and a momentum eigenstate.  (This is not to say that the
state of a system, as defined by its preparation, is ever
simultaneously an eigenstate of position and an eigenstate of
momentum.  The characterization of a state reduction necessarily
depends on the future of the system as well as its past.)

	Simple laboratory experiments provide examples of
physically equivalent beginning-to-end transitions.
Consider a single free particle of mass $m$ which is initially
measured to be at event $(x_{1},t_{1})$.  Suppose a later
measurement at time $t_{2}$ finds the particle at position
$x_{2}$.  This transition admits a classical description given by
the straight worldline joining the events $(x_{1},t_{1})$ and
$(x_{2},t_{2})$.  Some other transition would occur if the
particle were subject to a final measurement of its momentum
instead of its position.  However, the same classical description
would apply if the particle's momentum were found to have the
value $m
\frac{x_{2}-x_{1}}{t_{2}-t_{1}}$, and in that case both
transitions have the same probability.  Similarly, in a
scattering experiment, one may choose to measure either outgoing
particle positions or outgoing particle momenta.  The set of
allowed transitions depends on that choice, but the set of
stationary-phase paths may not.  Each transition of
non-negligible amplitude in one set can determine classical
particle worldlines which identify it with a physically
equivalent transition in the other set.  As in the toy cosmology,
each accessible final classical phase space trajectory identifies
multiple beginning-to-end transitions which are physically
equivalent.

\section{Quantum history cascades and Schroedinger's cat}

	It is here conjectured that realistic cosmology is in
some ways similar to the toy cosmology.  In the presence of
long-range interactions, Feynman paths are not simply particle
networks but also include bound states and classical fields.  A
cascade may still be defined as a class of stationary-phase
Feynman paths which share a common asymptotic trajectory in the
classical phase space.  The conjecture is that the big bang
initial state has special properties in common with the single
massive particle initial state in the toy cosmology.  In
particular, there may be a class of ``physically reasonable''
final bases, each of which has a subset which is in almost 1-to-1
correspondence with the set of cascades.  The set of cascades
then inherits a probability distribution from its correspondence
with any of those bases.  One may picture the universe as a
cascade sampled from that probability distribution.

	This picture is not intrinsic to the formalism of quantum
theory, however.  In either the toy cosmology or realistic
cosmology, there must be an operative constraint which is
responsible for the universe having a semi-classical description.
The formalism of quantum theory allows any Hilbert space state to
be the final state, so it does not exclude grotesque
beginning-to-end transitions which defy any semi-classical
description.  The transition to a generic final state does not
correspond to a cascade.  Indeed, if $\Psi_1$ and $\Psi_2$ are
physically reasonable final states which separately do correspond
to different cascades, then any non-trivial superposition of them
would define a beginning-to-end transition whose amplitude would
be calculated using paths with both asymptotic classical
trajectories.  The beginning-to-end transition to a non-trivial
superposition of $\Psi_1$ and $\Psi_2$ would not correspond to a
single asymptotic trajectory, and so it would not be a cascade.
In this sense, a non-trivial superposition of cascades cannot
itself be a cascade.  A generic beginning-to-end transition does
not admit a unique semi-classical description even
asymptotically, but is a grotesque superposition of physically
interpretable histories.

	A simple hypothesis for the operative constraint is that
the universe is sampled from a probability distribution over the
set of cascades.  This might seem to be inappropriate as a {\em
quantum} condition.  After all, it specifies that the end of the
universe corresponds to a {\em classical} state.  Moreover, it
cannot be formulated simply as a final boundary condition because
each cascade depends also on the initial state.  However, in
conjunction with a particular special initial condition (like the
single massive particle in the toy cosmology or the big bang
initial state in realistic cosmology), this hypothesis can be
equivalent to a class of reasonable final quantum constraints.
In the toy cosmology, for example, the final quantum constraint
could be that the final state must be a momentum eigenstate.
Alternatively, the constraint might require the final state to be
a position eigenstate, or a basis state of some other physically
reasonable basis.  Those different constraints have equivalent
effects.  In particular, transitions to the allowed and
accessible final states correspond to cascades.

	It is therefore plausible to maintain that a final
constraint requires the final state to be sampled from a special
set of basis states, exactly like the effect of a final
measurement in laboratory physics.  An allowed quantum history is
simply a beginning-to-end transition from the fixed initial state
to one of those final basis states.  Due to the remarkable nature
of the initial state, however, sampling an allowed transition may
correspond to sampling a cascade with its unique classical
description at asymptotically late times.  Moreover, many
different final bases may lead to the same probability
distribution over cascades.  Any of those final constraints, in
conjunction with the highly special initial state, results in the
universe being sampled from a probability distribution over the
various cascades.

	Consider once again cosmology in which a Schroedinger cat
experiment occurs.  An allowed history is now assumed to
correspond to a cascade.  It would be remarkable in the extreme
for a stationary-phase (semi-classical) path which describes a
dead cat at time $t$ to be asymptotically identical to another
stationary-phase path which describes a live cat at time $t$.
Unless circumstances are cleverly contrived, macroscopically
disparate semi-classical paths do not become identical in the
asymptotic future.  (In the toy cosmology, an asymptotic
classical phase space trajectory is generally consistent with at
most one particle network starting from the special initial
state.  It would be exceedingly rare to find two macroscopically
distinguishable networks -- i.e.  distinguished by a set of more
than $10^{23}$ interacting particle worldlines that differ in the
two networks -- which start from the special initial state and are
identical in the asymptotic future.)  Each cascade has an
asymptotic classical description.  The superposition of two
such cascades would not itself be a cascade.  A live-cat path and
a dead-cat path do not belong to any single cascade
unless they miraculously become identical in the asymptotic
future.

\section{Quantum theory and spacetime}

	In the context of quantum gravity, the final constraint
should ensure that the universe has an asymptotic spacetime
geometry rather than ending in a grotesque superposition of
disparate spacetimes.  Without a final constraint, there is no
natural explanation for the apparent spacetime structure which is
the foundation of physics.  Spacetime structure at intermediate
times may be regarded as due to state reductions which are
implicit in a quantum history which starts from the big bang and
has a classical geometry in the asymptotic future.

	A quantum history need not imply a unique classical
spacetime, however.  A particular history may be
compatible with multiple intermediate spacetime geometries.  For
example, one can consider the diffraction of a Planck-mass
particle by a double-slit diffractor and the particle's detection
on a subsequent surface.  A Planck-mass particle has a wavelength
comparable to its Schwarzschild radius, so wavelike behavior can
be expected along with effects of spacetime curvature.  If no
measurement is made of which path the particle follows, an
interference of the possibilities must occur.  The initial and
final states are compatible with either classical trajectory
through the diffractor.  The interfering spacetime geometries are
disparate, with the spacetime curvature (and corresponding
stress-energy) concentrated along one trajectory in one spacetime
and along a distinguishable trajectory in the other spacetime.
There is not a unique implicit spacetime geometry if there is no
record in the final state as to which intermediate geometry
occurred.  It is inappropriate to use an evolving state vector's
expectation value of stress-energy in an effort to salvage a
unique intermediate spacetime.  There are disparate intermediate
spacetimes contributing to the single quantum history.

\section{The Aharonov-Bohm effect}

	The present approach regards an entire quantum history as
a single entity.  The initial conditions together with a final
constraint determine the set of allowed quantum histories.
Only one history occurs.  It is not necessary that a quantum
history be interpretable as a future-blind evolution from its
initial state.  This has an important bearing on the
Aharonov-Bohm effect \cite{r10} as well as the EPR phenomenon.

	The Aharonov-Bohm effect concerns an electron which has
two possible classical worldlines from event E to event F.  The
two possible paths interfere, and the interference phase
difference depends quantitatively on an imposed electromagnetic
field.  The two classical worldlines constitute a closed curve in
spacetime.  The presence of an electromagnetic field contributes
a phase difference proportional to the electromagnetic field
(2-form) integrated over any 2-surface having the closed
worldline curve as its boundary.  Since an electromagnetic field
is a closed 2-form, that integral is independent of the choice of
spanning surface.

	What can seem surprising is that the electromagnetic
field makes this contribution to the phase difference even if it
is identically zero everywhere near the two possible worldlines.
Arguing that the electron interference should not be affected by
any physical fields which vanish everywhere near the possible
paths, Aharonov and Bohm ascribed physical reality to the
electromagnetic potential (or at least to the equivalence class
generated by adding exact differentials).  Using Stokes' theorem,
the surface integral can be re-expressed as a line integral of a
potential 1-form over the boundary curve.  If the surface
integral is non-zero (so the electromagnetic contribution to the
phase difference is non-zero), then the potential 1-form cannot
vanish everywhere on the electron paths.

	In the present approach, the motivation for ascribing
physical reality to the potential 1-form is absent.  The entire
quantum history exists as a single entity.  The probability of a
given history may depend on the phase difference at event F, and
that is determined by the electromagnetic field in the causal
future of E and in the causal past of F.  It is not necessary
that the phase difference be determined exclusively by fields in
the neighborhood of the electron paths.  There is no reason to
prefer the line integral calculation over the surface integral.
The phase difference (and hence probabilities for various allowed
quantum histories) can be determined by the electromagnetic field
without reference to potentials.

\section{Discussion}

	EPR experiments do not pose any problem of interpretation
if one imagines that an entire quantum history is selected from a
set of allowed histories each of which satisfies all conservation
laws.  Unfortunately, an initial state, by itself, does not
define a suitable set of quantum histories.  In laboratory
experiments, the set of allowed quantum histories is determined
by the initial state together with a final constraint which
requires the final state to be sampled from the eigenstates of
the final measurement operator.  The proposal here is to regard
the quantum universe to be subject to an analogous final
constraint, even though there is no external apparatus to perform
a final measurement.  The nature of the final constraint has been
the subject of speculation in sections 3-5.

	This approach to quantum theory avoids interpretation
problems without altering the theory itself.  In particular, it
is not necessary to introduce a mechanism like von Neumann's
R-process \cite{s2} to account for state reductions.  There is no
modification of the formalism, so there can be no conflict with
any experimental result.

	The point of view advocated here seems to be shared, in
part, by other authors.  Griffiths \cite{g1}, Gell-Mann and
Hartle \cite{r12}, and Omnes \cite{r13} have emphasized the use
of quantum histories (although they ascribe a different meaning
to the term ``quantum
history'').  Schulman \cite{r14}, Costa de
Beauregard \cite{r15}, Cramer \cite{r16}, and others have
recognized the relevance of future conditions for quantum
systems.  Future conditions are incompatible with the strong
principle of causality.  As argued in section 3, however, the
type of future constraint suggested here does not lead to
faster-than-light signal transmission.  Moreover, many authors
have concluded that the strong causality principle is
incompatible with quantum theory.  Even Bohr \cite{r17} emphasized that
quantum theory ``entails the necessity of a final renunciation of
the classical ideal of causality.''

	{\em Why} there should be a final constraint on the
universe may be a metaphysical question.  Without such a boundary
condition, however, quantum theory poses formidable
interpretation problems.  The universe is a quantum history with
a semi-classical description.  Should that be regarded as a
fluke, or due to some additional structure which must be
incorporated in quantum theory itself, or simply the consequence
of an operative boundary condition?  Occam's razor favors the
boundary condition.  A final constraint can be invoked to explain
why quantum states get reduced to physically interpretable states
of Hilbert space.  It puts the quantum into quantum theory.

	The picture presented here of the final state being
sampled from a probability distribution over a set of allowed
final states can only be a conjecture.  It could be, for example,
that boundary conditions on the universe include both its
definite initial state $\Psi_{0}$ and also its definite final
state $\Psi_{f}$.  Even if one could survey the universe from
start to finish, it should not be possible to deduce whether the
actual final state is sampled from a probability distribution or
is dictated uniquely as a boundary condition.  A dictated final
state might seem like boring physics since the quantum history
would then be fully specified by the imposed boundary conditions.
That would be a return to the classical situation where (initial)
boundary conditions determine the entire history uniquely.  The
approach proposed here allows the usual quantum indeterminacy in
deriving the actual quantum history from initial conditions.
Unless it becomes apparent that the universe is developing toward
some highly special final state, there is no compelling reason to
suppose that a boundary condition dictates a unique final state.
There would be no elegance in some undistinguished final state being
imposed as the only allowed final state.  Highly specific {\em
initial} conditions apparently do pertain to the universe.  There
is no other natural way to account for the low entropy at early
cosmological times.  The inference of a final constraint is
similar.  No other natural explanation has emerged to account for
the universe having a reasonable semi-classical description
rather than a grotesque evolution.

	Although the entire quantum history is here presumed to
have been selected as a single entity with a unique future, the
future is unpredictable because there are an infinite number of
allowed histories which include the universe's sequence of state
reductions up to the present time.  It is not possible to
identify which of the allowed histories is the unique actual
complete history.  A twin pair of EPR particles may now be
in spin eigenstates corresponding to a reference axis to be
chosen tomorrow.  Their present spin eigenstates are influenced
by the future measurements.  Those spin states cannot be used to
predict the future measurement axis, however, since a spin
measurement on either of them today would negate the influence of
tomorrow's measurements.

	Essential to this approach is the inference, based on
consideration of the EPR phenomenon, that the entire
quantum history (through the end of time) has been selected as a
single entity.  Though the future is not
predictable, there is a unique future as well as a unique past.
God does not continue to shoot dice.  The dice were rolled once
to select the entire quantum universe.


\begin{thebibliography}{999}
\bibitem{s1}DeWitt, B.S. and Graham, N. (eds), {\em The Many-Worlds
Interpretation of Quantum Mechanics}, Princeton Univ., Princeton,
NJ (1973); DeWitt, B.S., {\em Physics Today} {\bf 23}(9), 30 (1970).
\bibitem{s2}von Neumann, J., {\em Mathematical Foundations of
Quantum Mechanics}, Princeton University Press (Princeton, 1932).
\bibitem{r2}Einstein, A., Podolsky, B., and Rosen, N., {\em Phys.
Rev.} {\bf 47}, 777 (1935).
\bibitem{r3}Bohm, D., {\em Quantum Theory}, Prentice-Hall,
Englewood, NJ, 614 (1951).
\bibitem{r4}Aspect, A., Dalibard, J., and Roger, G.,
{\em Phys. Rev. Lett.} {\bf 49}, 1804 (1982).
\bibitem{r5}Bell, J.S., {\em Physics} {\bf 1}, 195 (1964).
\bibitem{r6}Mermin, N.D., {\em Physics Today} {\bf 38}(4), 38 (1985).
\bibitem{g1}Griffiths, R.B., {\em J. Stat. Phys.} {\bf 36} 219 (1984).
\bibitem{r7}Wigner, E.P.,  {\em Am. J. Phys.} {\bf 31}, 6 (1963).
\bibitem{r8}Schroedinger, E., {\em Proc. Cambridge Philos. Soc.} {\bf 31},
555 (1935).
\bibitem{g2}Feynman, R.P. and Hibbs, A.R. {\em Quantum Mechanics
and Path Integrals}, McGraw-Hill (1965).
\bibitem{r10}Aharonov, Y. and Bohm, D., {\em Phys. Rev.} {\bf
115}, 485 (1959).
\bibitem{r12}Gell-Mann, M. and Hartle, J.B., in {\em Complexity,
Entropy and the physics of Information}, ed: W. Zurek
(Addison-Wesley, Redwood City, CA, 1991).
\bibitem{r13}Omnes, R., {\em Rev. Mod. Phys.} {\bf 64}, 339
(1992)
\bibitem{r14}Schulman, L.S., {\em Ann. Phys.} {\bf 212}, 315
(1991)
\bibitem{r15}Costa de Beauregard, O., {\em Found. Phys.} {\bf
15}, 871 (1985)
\bibitem{r16}Cramer, J.G., {\em Rev. Mod. Phys.} {\bf 58}, 647 (1986)
\bibitem{r17}Bohr, N., {\em Phys. Rev.} {\bf 48}, 696 (1935)
\end{thebibliography}
\end{document}